\documentclass[superscriptaddress,amsmath, nofootinbib]{revtex4}
\usepackage{amsfonts}
\usepackage{graphicx}
\usepackage[latin1]{inputenc}
\usepackage{amsmath}
\usepackage{amssymb}

\begin{document}


\title{Lagrangian formulation for Mathisson-Papapetrou-Tulczyjew-Dixon (MPTD) equations}
\author{Walberto Guzm\'an Ram\'irez }
\email{wguzman@cbpf.br} \affiliation{Depto. de Matem\'atica, ICE, Universidade Federal de Juiz de Fora, MG, Brazil}

\author{Alexei A. Deriglazov }
\email{alexei.deriglazov@ufjf.edu.br} \affiliation{Depto. de Matem\'atica, ICE, Universidade Federal de Juiz de Fora,
MG, Brazil} \affiliation{Laboratory of Mathematical Physics, Tomsk Polytechnic University, 634050 Tomsk, Lenin Ave. 30,
Russian Federation}

\date{\today}

\begin{abstract}
We obtain Mathisson-Papapetrou-Tulczyjew-Dixon equations of a rotating body with given values of spin and momentum
starting from Lagrangian action without auxiliary variables. MPTD-equations correspond to minimal interaction of our
spinning particle with gravity. We shortly discuss some novel properties deduced from the Lagrangian form of
MPTD-equations: emergence of an effective metric instead of the original one, non-commutativity of coordinates of
representative point of the body, spin corrections to Newton potential due to the effective metric as well as spin
corrections to the expression for integrals of motion of a given isometry.
\end{abstract}

\maketitle 

\section{Introduction}

The description of spinning bodies in general relativity is an old subject, which is under intensive study over the
last 70 years. The first results concerning equations of motion of a test body in a given background were reported by
Mathisson \cite{Mathisson:1937zz} and Papapetrou \cite{Papapetrou:1951pa}. They assumed that the structure of test body
can be described by a set of multipoles and have taken the approximation which involves only first two terms (the
pole-dipole approximation). The equations are then derived by integration of conservation law for the energy-momentum
tensor, $T^{\mu\nu}{}_{;\mu}=0$. A manifestly covariant formulation was given by Tulczyjew \cite{Tulc} and Dixon
\cite{Dixon1964} and is under detailed investigation by many groups. In this work we will refer the equations
(6.31)-(6.33) in \cite{Dixon1964} as MPTD-equations. Detailed analysis and interpretation of these equations and their
generalizations \cite{Trautman2002, Costa2012, Costa2014, Khriplovich1989, KyrianSemerak, Fenyk2013, Hojman3,
Poplawski:2010, Burinskii2014, Mohseni2015, Holten2015, Kunst2015, Bini2014, Singh2002, Singh2003} is an actual task
since they are widely used now to account spin effects in compact binaries and rotating black holes, see
\cite{Pomeranskii, Porto2011, Han2010, Khriplovich99, Khriplovich96, Pomeranskii1998} and references therein.

It should be interesting to obtain these equations starting from an appropriate Lagrangian action. The vector models of
spin yields one possible way to attack the problem. In these models, the basic variables in spin-sector are
non-Grassmann vector $\omega^\mu$ and its conjugated  momentum $\pi_\mu$. The spin-tensor is a composed quantity
constructed from these variables, $S^{\mu\nu}=2(\omega^\mu\pi^\nu-\omega^\nu\pi^\mu)$. To have a theory with right
number of physical degrees of freedom for the spin (two for an elementary particle with spin one-half, and three for a
rotating body in the pole-dipole approximation), some constraints on the eight basic variables must be imposed. This is
the main difficulty: besides the equations of motion, the variational problem should produce these constraints. Even
for the free theory in flat space, this turns out to be rather non trivial problem \cite{hanson1974, mukunda1982,
Bailey, Wojciech, Gerald}. Here we propose the Lagrangian action without auxiliary variables which, besides the
equations of motion, yields all the desired constrains.

To point out some advantages of the vector model, let us compare it with the approach developed in \cite{barausse2009}
for the description of relativistic top \cite{hanson1974} in curved background. First, in the vector model we have four
basic variables in spin-sector instead of six (called $\phi_a$ in \cite{barausse2009}) for the top. Taking into account
that we present the Lagrangian without auxiliary variables, the vector model yields more economic formulation. Second,
our primary constraints (see $T_6$ and $T_7$ below) follow from the variational problem and yield the spin
supplementary condition (\ref{condition1}). In the work \cite{barausse2009} the condition has been added by hand and
then considered as a first-class constraint of the formulation. This implies, in particular, that all the basic
variables (including the position variables) of the theory \cite{barausse2009} are unobservable quantities. Third, the
vector model yields two physical degrees of freedom in spin-sector. Hence it can be used for the description of both a
rotating body (see below), and for an elementary particle with spin. In particular, canonical quantization of the
vector model has been considered in \cite{DPM2}.

The work is organized as follows. In Section 2 we present Lagrangian action without auxiliary
variables\footnote{Variational problem with four auxiliary variables has been constructed in \cite{DPW2}.} for our
spinning particle in an arbitrary curved background, and obtain its Hamiltonian formulation. Section 3 contains
detailed derivation and analysis of both Lagrangian and Hamiltonian equations. The particle has fixed value of spin and
two physical degrees of freedom in the spin-sector. We also present a modification which yields the model of
Hanson-Regge type \cite{hanson1974}, with unfixed value of spin and four physical degrees of freedom. In Section 4 we
present the MPTD-equations in the form convenient for our analysis. Here we follow the ideas of Dixon \cite{Dixon1964}
and add the mass-shell condition to MPTD-equations, transforming them into the Hamiltonian system. This allows us to
compare MPTD-equations with those of Section 3. We show that the class of trajectories of MPTD-equations with any given
values of integration constants (squares of spin and of momentum) is described by our spinning particle with properly
chosen mass and spin. In section 5 we discuss some novel properties which can be immediately deduced from the
Lagrangian form of MPTD-equations.

{\bf Notation.} The dynamical variables are taken in arbitrary parametrization $\tau$, then $\dot
x^\mu=\frac{dx^\mu}{d\tau}$, $\mu, \nu=0, 1, 2, 3$. Covariant derivative is $\nabla P^\mu=\frac{dP^\mu}{d\tau}
+\Gamma^\mu_{\alpha\beta}\dot x^\alpha P^\beta$ and curvature is
$R^\sigma{}_{\lambda\mu\nu}=\partial_\mu\Gamma^\sigma{}_{\lambda\nu} -\partial_\nu
\Gamma^\sigma{}_{\lambda\mu}+\Gamma^\sigma{}_{\beta\mu}\Gamma^{\beta}{}_{\lambda\nu}-
\Gamma^\sigma{}_{\beta\nu}\Gamma^{\beta}{}_{\lambda\mu}$. The square brackets mean antisymmetrization,
$\omega^{[\mu}\pi^{\nu]}=\omega^\mu\pi^\nu-\omega^\nu\pi^\mu$. We use the condensed notation $\dot x^\mu G_{\mu\nu}\dot
x^\nu=\dot xG\dot x$,  $N^\mu{}_\nu\dot x^\nu=(N\dot x)^\mu$, $\omega^2=g_{\mu\nu}\omega^\mu\omega^\nu$, and so on.
Notation for the scalar functions constructed from second-rank tensors are $\theta S= \theta^{\mu\nu}S_{\mu\nu}$,
$S^2=S^{\mu\nu}S_{\mu\nu}$.

\section{Lagrangian and Hamiltonian formulations \label{curvedS}}
The variational problem for vector model of spin interacting with electromagnetic and gravitational fields can be
formulated with various sets of auxiliary variables \cite{DPW2, DPM2, DPM3, deriglazov2014Monster}. For the free theory
in flat space there is the Lagrangian action without auxiliary variables. Configuration space consist of the position
$x^\mu(\tau)$ and the vector $\omega^\mu(\tau)$ attached to the point $x^\mu$. The action reads
\begin{eqnarray}\label{Lfree}
S=-\frac{1}{\sqrt{2}} \int d\tau \sqrt{m^2c^2 -\frac{\alpha}{\omega^2}} ~  \sqrt{-\dot x N \dot x - \dot\omega N
\dot\omega +\sqrt{[\dot x N\dot x + \dot\omega N \dot\omega]^2- 4 (\dot x N \dot\omega )^2}}.
\end{eqnarray}
The matrix $N_{\mu\nu}$ is the projector on the plane orthogonal to $\omega^\nu$
\begin{equation}\label{projector}
N_{\mu\nu}= \eta_{\mu\nu}-\frac{\omega_\mu \omega_\nu}{\omega^2}, \quad \mbox{then}  \quad N_{\mu\nu} \omega^\nu=0.
\end{equation}
Below we use the notation
\begin{equation}\label{T121}
T\equiv [\dot x N\dot x + \dot\omega N \dot\omega]^2- 4 (\dot x N \dot\omega )^2 \, .
\end{equation}
The double square-root structure in the expression (\ref{Lfree}) seem to be typical for the vector models of spin
\cite{hanson1974}. The Lagrangian depends on one free parameter $\alpha$ which determines the value of spin. The value
$\alpha=\frac{3\hbar^2}{4}$ corresponds to a spin one-half particle. In the spinless limit, $\alpha=0$ and
$\omega^\mu=0$, the equation (\ref{Lfree}) reduces to the standard expression, $-mc\sqrt{-\dot x^\mu\dot x_\mu}$. The
equivalent Lagrangian with one auxiliary variable $\lambda(\tau)$ is
\begin{eqnarray}\label{FF.1}
L=\frac{1}{4\lambda}\left[\dot xN\dot x+\dot\omega N\dot\omega-T^{\frac12}\right]-\frac{\lambda}{2}(m^2c^2-\frac{\alpha}{\omega^2}).
\end{eqnarray}
Switching off the spin variables $\omega^\mu$ from Eq. (\ref{FF.1}), we arrive at familiar Lagrangian of spinless
particle $L=\frac{1}{2\lambda}\dot x^2-\frac{\lambda}{2}m^2c^2$. In this formulation the model admits interaction with
an arbitrary electromagnetic field. The interacting theory is obtained \cite{deriglazov2014Monster} adding the minimal
interaction term, $\frac{e}{c}A_\mu\dot x^\mu$, and replacing $\dot\omega^\mu$ by
$D\omega^\mu\equiv\dot\omega^\mu-\lambda\frac{e\mu}{c}F^{\mu\nu}\omega_\nu$, where $\mu$ is the magnetic moment.

The Frenkel spin-tensor \cite{Frenkel} in our model is a composite quantity constructed from $\omega^\mu$ and its
conjugated momentum $\pi^\mu=\frac{\partial L}{\partial\dot\omega_\mu}$ as follows:
\begin{eqnarray}\label{FF.2}
S^{\mu\nu}=2(\omega^\mu\pi^\nu-\omega^\nu\pi^\mu)=(S^{i0}=D^i, ~ S_{ij}=2\epsilon_{ijk}S_k).
\end{eqnarray}
Here $S_i$ is three-dimensional spin-vector and $D_i$ is dipole electric moment \cite{ba1}. The model is invariant
under reparametrizations and local spin-plane symmetries \cite{DPM1}. The latter symmetry acts on $\omega^{\mu}$ and $\pi^{\mu}$
but leaves $S^{\mu\nu}$ invariant. So only $S^{\mu\nu}$ is an observable quantity. In their work \cite{hanson1974}, Hanson and
Regge analyzed whether the spin-tensor interacts directly with an electromagnetic field, and concluded on impossibility
to construct the interaction in closed form. In our model an electromagnetic field interacts with $\omega^\mu$ from which the
spin-tensor is constructed.

The minimal interaction with gravitational field can be achieved by covariantization of the formulation (\ref{Lfree}).
In the expressions (\ref{Lfree})-(\ref{T121}) we replace $\eta_{\mu\nu}\rightarrow g_{\mu\nu}$, and usual derivative by
the covariant one, $\dot\omega^\mu\rightarrow \nabla\omega^\mu=\frac{d\omega^\mu}{d\tau} +\Gamma^\mu_{\alpha\beta}\dot
x^\alpha\omega^\beta$. Thus our Lagrangian in a curved background reads
\begin{eqnarray}\label{L-curved}
L =-\frac{1}{\sqrt{2}} \left[ m^2c^2 -\frac{\alpha}{\omega^2} \right]^{\frac 12} ~  \sqrt{-\dot x N \dot x -
\nabla\omega N \nabla\omega +\sqrt{[\dot x N\dot x + \nabla\omega N \nabla\omega]^2- 4 (\dot x N \nabla\omega )^2}} \cr
\equiv -\frac{1}{\sqrt{2}} \left[ m^2c^2 -\frac{\alpha}{\omega^2} \right]^{\frac 12} ~ L_0.
\qquad \qquad \qquad
\qquad  \qquad \qquad \qquad \qquad \qquad  \qquad \qquad \qquad
\end{eqnarray}
Velocities $\dot x^\mu$, $\nabla\omega^\mu$ and projector $N_{\mu\nu}$ transform like contravariant vectors and covariant
tensor, so the action is manifestly invariant under general-coordinate transformations.

Let us construct Hamiltonian formulation of the model (\ref{L-curved}). Conjugate momenta for $x^\mu$ and $\omega^\mu$
are $p_\mu=\frac{\partial L}{\partial\dot x^\mu}$ and  $\pi_\mu=\frac{\partial L}{\partial\dot\omega^\mu}$
respectively. Due to the presence of Christoffel symbols in $\nabla\omega^\mu$, the conjugated momentum $p_\mu$ does not transform as a
vector, so it is convenient to introduce the canonical momentum
\begin{equation}
P_\mu\equiv p_\mu-\Gamma^\beta_{\alpha\mu}\omega^\alpha\pi_\beta \, ,
\end{equation}
the latter  transforms as a vector under general transformations of coordinates. Manifest form of the momenta is as follows:
\begin{eqnarray}\label{p1}
P_\mu& =& \frac{1}{\sqrt{2}L_0}\left[m^2c^2 -\frac{\alpha}{\omega^2}\right]^{\frac12}\left[ N_{\mu\nu} \dot x^\nu - K_\mu \right]
\end{eqnarray}
\begin{eqnarray}\label{p2}
\pi_\mu &=& \frac{1}{\sqrt{2}L_0}\left[m^2c^2 -\frac{\alpha}{\omega^2}\right]^{\frac12} \left[ N_{\mu\nu} \nabla\omega^\nu - R_\mu \right] \, ,
\end{eqnarray}
with
\begin{eqnarray}
K_\mu=T^{-1/2} \left[ (\dot x N \dot x +  \nabla\omega N \nabla\omega) (N \dot x)_\mu -
2 (\dot x N \nabla\omega) (N \nabla\omega)_\mu \right] \, , \nonumber \\
R_\mu=T^{-1/2} \left[ (\dot x N \dot x +  \nabla\omega N \nabla\omega) (N \nabla\omega)_\mu -
2 (\dot x N \nabla\omega) (N \dot x)_\mu \right] \, . \nonumber
\end{eqnarray}
These vectors  obey the following algebraic  identities
\begin{eqnarray}\label{rel-1}
K^2= \dot x N\dot x \, , \quad R^2=\nabla \omega N \nabla \omega \, , \quad KR = -\dot xN\nabla \omega \, , \quad
\dot x R +\nabla \omega K =0 , \quad \quad K\dot x  + R \nabla \omega = T^{\frac 12} \, .
\end{eqnarray}
Using  (\ref{projector})  we  conclude that   $\omega\pi=0$ and $P\omega =0$, that is we found two primary constraints.
Using the relations  (\ref{rel-1}) we find one more primary constraint   $P\pi =0$. Besides, computing $P^2+\pi^2$
given by (\ref{p1}) and (\ref{p2}) we see that all the terms with derivatives vanish, and we obtain the last primary
constraint
\begin{equation}\label{p3}
T_1\equiv P^2 + m^2 c^2 +  \pi^2 - \frac{\alpha}{\omega^2}=0 \, .
\end{equation}
In the result, the action (\ref{L-curved}) implies four primary constraints,  $T_1$ and
\begin{eqnarray}\label{primary}
T_5\equiv\omega\pi=0, \quad T_6\equiv P\omega =0 , \quad
T_7\equiv P\pi =0\,.
\end{eqnarray}
The  Hamiltonian is constructed excluding velocities from the expression
\begin{equation}\label{Hamiltonian-0}
H= p_\mu \dot x +\pi \dot\omega - L + \lambda_i T_i\equiv P\dot x + \pi\nabla\omega - L + \lambda_i T_i\, ,
\end{equation}
where $\lambda_i$ ($i=1,5,6,7$) are the  Lagrangian multipliers associated with the primary constraints. From
(\ref{p1}) and (\ref{p2}), we observe the equalities $ P\dot x= (\sqrt{2}L_0)^{-1} (m^2c^2-
\frac{\alpha}{\omega^2})^{\frac 12} [\dot x N \dot x - \dot x K]$ and  $ \pi\nabla\omega =
(\sqrt{2}L_0)^{-1}(m^2c^2-\frac{\alpha}{\omega^2})^{\frac 12} [\nabla\omega N\nabla\omega  - \nabla\omega R]$. Together
with (\ref{rel-1})  they imply
$P\dot x + \pi\nabla\omega =L$.
Using this in (\ref{Hamiltonian-0}), we conclude that the Hamiltonian is composed from the primary constraints
\begin{equation}\label{Hamiltonian1}
H=\frac{\lambda_1}{2}\left( P^2 + m^2 c^2 + \pi^2 - \frac{\alpha}{\omega^2} \right) + \lambda_5 (\omega\pi) +\lambda_6
(P\omega) +\lambda_7 (P\pi )\, .
\end{equation}
The full set of phase-space coordinates consists of the pairs $x^\mu, p_\mu$ and  $\omega^\mu, \pi_\mu$.  They fulfill
the fundamental Poisson brackets $\{x^\mu ,  p_\nu\}=\delta^\mu_{\nu}$ and $\{\omega^\mu , \pi_\nu\}=\delta^\mu_\nu$, then
$\quad \{P_\mu , P_\nu \} = R^\sigma_{\ \lambda \mu\nu} \pi_\sigma \omega^\lambda$, $\{P_\mu , \omega^\nu
\}=\Gamma^\nu_{\mu\alpha}\omega^\alpha$, $\{ P_\mu , \pi_\nu \}=- \Gamma^\alpha_{\mu\nu} \pi_\alpha$. For the
quantities $x^\mu$, $P^\mu$ and $S^{\mu\nu}$ these brackets imply the typical relations used by people for spinning
particles in Hamiltonian formalism
\begin{eqnarray}\label{br}
\{ x^\mu , P_\nu \} =\delta^\mu_\nu,  \quad \{ P_\mu , P_\nu\}=-\frac14 R_{\mu\nu\alpha\beta}S^{\alpha\beta}, \quad \{
P_\mu, S^{\alpha\beta} \}=\Gamma^{\alpha}_{\mu\sigma}S^{\sigma\beta}-\Gamma^\beta_{\mu\sigma}S^{\sigma\alpha} \, , \cr
\{S^{\mu\nu},S^{\alpha\beta}\}= 2(g^{\mu\alpha} S^{\nu\beta}-g^{\mu\beta} S^{\nu\alpha}-g^{\nu\alpha} S^{\mu\beta}
+g^{\nu\beta} S^{\mu\alpha})\,. \qquad \qquad \quad
\end{eqnarray}
To reveal the higher-stage constraints and the Lagrangian multipliers we study the equations $\dot T_i=\{ T_i , H\}=0$. $T_5$ implies the
secondary constraint
\begin{equation}\label{secondary}
\dot T_5=0 \quad \Rightarrow \quad T_3\equiv \pi^2 - \frac{\alpha}{\omega^2}\approx 0,
\end{equation}
then  $T_1$ can be replaced on $P^2 + m^2c^2 \approx 0$. Preservation in time of $T_7$ and $T_6$ gives the Lagrangian
multipliers $\lambda_6$ and $\lambda_7$
\begin{equation}\label{FF3}
\lambda_6 = \frac{\lambda_1 R_{(\pi)}}{2M^2c^2}, \quad \lambda_7= -\frac{\lambda_1 R_{(\omega)}}{2M^2c^2} \, ,
\end{equation}
where we have denoted
\begin{eqnarray}\label{Rpi}
R_{(\pi)}= 2R_{\alpha\beta\mu\nu} \omega^\alpha \pi^\beta \pi^\mu P^\nu \, ,   \qquad
R_{(\omega)}= 2R_{\alpha\beta\mu\nu} \omega^\alpha \pi^\beta \omega^\mu P^\nu \,.
\end{eqnarray}
\begin{equation}
M^2=m^2+\frac{1}{c^2} R_{\alpha \mu\beta\nu}\omega^\alpha\pi^\mu\omega^\beta\pi^\nu \equiv m^2+\frac{1}{16c^2}\theta S
\,,
\end{equation}
\begin{equation}\label{theta}
\theta_{\mu\nu} \equiv R_{\alpha\beta\mu\nu}S^{\alpha\beta} \, .
\end{equation}

Preservation in time of $T_1$ gives the equation $\lambda_6 R_{(\omega)} + \lambda_7 R_{(\pi)} =0 $ which is
identically satisfied by virtue of (\ref{FF3}).  No more constraints are generated after this step . We summarize the
algebra of Poisson between the constraints in the Table \ref{algebra-constraints}. $T_6$ and $T_7$ represent a pair of
second-class constraints, while $T_3$, $T_5$ and the  combination
\begin{equation} \label{T0}
T_0 = T_1 + \frac{R_{(\pi)}}{M^2c^2} T_6 - \frac{R_{(\omega)}}{M^2c^2} T_7 \, ,
\end{equation}
are the first-class constraints. The presence of two primary first-class constraints $T_5$ and $T_0$ is in
correspondence with the fact that two Lagrangian multipliers remain undetermined. This also is in agreement with the
invariance of our action with respect to two local symmetries mentioned above. Taking into account that each
second-class constraint rules out one phase-space variable, whereas each first-class constraint rules out two
variables, we have the right number of spin degrees of freedom, $8-(2+4)=2$.

We point out that the first-class constraint $T_3=\pi^2 - \frac{\alpha}{\omega^2}\approx 0$ can be replaced on the pair
\begin{eqnarray}\label{primary0}
\pi^2=\mbox{const}, \qquad \omega^2=\mbox{const},
\end{eqnarray}
this gives an equivalent formulation of the model. The Lagrangian which implies the constraints (\ref{primary}) and
(\ref{primary0}) has been studied in \cite{DPW2, DPM2, DPM3, Rempel2015}. Hamiltonian and Lagrangian equations for
physical variables of the two formulations coincide \cite{deriglazov2014Monster}, which proves their equivalence.

Using (\ref{FF3}), we can present the Hamiltonian (\ref{Hamiltonian1}) in the form
\begin{eqnarray}\label{Hamiltonian}
H= \frac{\lambda_1}{2} \left( P^2 + m^2c^2 +  \frac{R_{(\pi)}   (P\omega) -  R_{(\omega)}(P\pi )}{M^2c^2}
\right)+\frac{\lambda_1}{2}\left(  \pi^2 - \frac{\alpha}{\omega^2} \right)  + \lambda_5 (\omega\pi) \, .
\end{eqnarray}

\begin{table}
\caption{Algebra of constraints} \label{algebra-constraints}
\begin{center}
\begin{tabular}{|c|c|c|c|c|c|}
\hline
                              & $\qquad T_1 \qquad$  & $T_3$         & $T_5$          & $T_6$                 & $T_7$       \\  \hline \hline
$T_1=P^2+m^2 c^2$         & 0             & 0             & 0              &$ R_{(\omega)}$                    &$ R_{(\pi)}$     \\

                              &               &               &           &            &              \\ \hline
$ T_3=\pi^2 -\frac{\alpha}{\omega^2}    $               & 0             &   0   &  $-2T_3$             & $-2T_7$   & $-2T_6/\omega^2$   \\
& ${}$ &      &           &                &      \\
\hline
$T_5=\omega\pi$            & 0             & 0  & $0$            & $-T_6$          & $T_7$     \\
& ${}$ &      &           &                &        \\   \hline
$T_6=P\omega$               & $-R_{(\omega)}$    & $2T_7$      & $T_6$   &     0                 & $-M^2c^2$  \\
& ${}$ &      &           &                &        \\  \hline
$T_7=P \pi$       & $-R_{(\pi)}$  &$2T_6/\omega^2$  & $-T_7$&   $M^2c^2$               & 0        \\
                              & ${}$ &      &           &                &     \\   \hline
\end{tabular}
\end{center}
\end{table}
%

%

%
\section{Equations of motion\label{EquationsofMotion}}
The dynamics of basic variables is governed by Hamiltonian equations $\dot z=\{z , H\}$, where $z=(x, p, \omega, \pi)$, and
the Hamiltonian is given in (\ref{Hamiltonian}). The equations can be written in a manifestly covariant form as follows:
\begin{eqnarray}
\dot x^\mu= \lambda_1 \left[ P^\mu +  (2M^2c^2)^{-1}(R_{(\pi)} \omega^\mu-R_{(\omega)}\pi^\mu) \right] \, ,
\qquad \qquad \qquad  \label{motion-x-1} \\
\nabla P_\mu =R^\alpha_{\ \ \beta\mu\nu} \pi_\alpha \omega^\beta  \dot x^\nu \, , \qquad \qquad \qquad
\qquad \qquad \qquad \qquad \qquad \label{motion-P}  \\
\nabla\omega^\mu =- \lambda_1\frac{R_{(\omega)}}{2M^2c^2} P^\mu + \lambda_5 \omega^\mu +\lambda_1\pi^\mu \, , \quad
\nabla\pi_\mu= -\lambda_1\frac{R_{(\pi)}}{2M^2c^2} P_\mu- \lambda_5 \pi_\mu -\lambda_1\frac{\omega_\mu}{\omega^2} \, .
\label{motion-pi-omega}
\end{eqnarray}
Neither constraints nor equations of motion do not determine the functions $\lambda_1$ and $\lambda_5$. Their presence
in the equations of motion implies that evolution of our basic variables is ambiguous. This is in correspondence with
two local symmetries presented in the model. According to general theory \cite{13, gitman1990quantization,
deriglazov2010classical}, variables with ambiguous dynamics do not represent observable quantities, so we need to
search  for variables that can be candidates for observables. Consider  the  antisymmetric tensor (\ref{FF.2}).
%
%
Aa a consequence of $T_6=0$ and $T_7=0$, this obeys the Tulczyjew supplementary condition
\begin{equation}
S^{\mu\nu}P_\nu = 0 \, .\label{condition1}
\end{equation}
Besides, the constraints $T_3$ and $T_5$ fix the value of square
\begin{equation}
S^{\mu\nu} S_{\mu\nu} = 8\alpha, \label{condition2}
\end{equation}
so we identify $S^{\mu\nu}$ with the Frenkel spin-tensor \cite{Frenkel}. The equations (\ref{condition1}) and
(\ref{condition2}) imply that only two components of spin-tensor are independent, as it should be for spin one-half
particle. Equations of motion for $S^{\mu\nu}$ follow from (\ref{motion-pi-omega}). Besides, using (\ref{Rpi}) we express
equations (\ref{motion-x-1}) and (\ref{motion-P}) in terms of the spin-tensor. This gives the system
\begin{eqnarray}
\dot x^\mu &=& \lambda_1\left[ P^\mu +\frac{1}{8M^2c^2} S^{\mu\beta}\theta_{\beta\alpha}P^\alpha\right] \, , \label{motion-x} \\
\nabla P_\mu &=&- \frac{1}{4} R_{\mu\nu\alpha\beta} S^{\alpha\beta}\dot  x^\nu\equiv-\frac{1}{4}\theta_{\mu\nu}\dot
x^\nu
\, ,  \label{motion-P-2.2} \\
\nabla S^{\mu\nu} &=& 2(P^\mu \dot x^\nu - P^\nu \dot x^\mu )\,, \label{motion-J.3}
\end{eqnarray}
where $\theta$ has been defined in (\ref{theta}). Eq. (\ref{motion-J.3}), contrary to the equations
(\ref{motion-pi-omega}) for $\omega$ and $\pi$, does not depend on $\lambda_5$. This proves that the spin-tensor is
invariant under local spin-plane symmetry. The remaining ambiguity due to  $\lambda_1$ is related with
reparametrization invariance and disappears when we work with physical dynamical variables $x^i(t)$. Equations
(\ref{motion-x})-(\ref{motion-J.3}) together with (\ref{condition1}) and (\ref{condition2}), form a closed system which
determines evolution of a spinning particle.

To obtain the Hamiltonian equations we can equally use the Dirac bracket constructed with help of second-class
constraints
\begin{equation}\label{DB}
\{A , B \}_D = \{A , B\} -\frac{1}{M^2c^2} \left[ \{ A , T_6 \} \{T_7 , B\} - \{ A , T_7\}\{ T_6 , B\} \right] \,.
\end{equation}
Since the Dirac bracket of a second-class constraint with any quantity vanishes, we can now omit $T_6$ and $T_7$ from
(\ref{Hamiltonian}), this yields the Hamiltonian
\begin{eqnarray}\label{Hamiltonian.0}
H_1= \frac{\lambda_1}{2} \left( P^2 + m^2c^2\right) +\frac{\lambda_1}{2}\left(  \pi^2 - \frac{\alpha}{\omega^2} \right)
+ \lambda_5 (\omega\pi) \,.
\end{eqnarray}
Then the equations (\ref{motion-x-1})-(\ref{motion-pi-omega}) can be obtained according the rule $\dot z=\{z, H_1\}_D$.
The quantities $x^\mu$, $P^\mu$ and $S^{\mu\nu}$, being invariant under spin-plane symmetry, have vanishing brackets
with the corresponding first-class constraints $T_3$ and $T_5$. So, obtaining equations for these quantities, we can
omit the last two terms in $H_1$, arriving at the familiar relativistic Hamiltonian
\begin{eqnarray}\label{Hamiltonian.0.02}
H_2= \frac{\lambda_1}{2} \left( P^2 + m^2c^2\right) \,.
\end{eqnarray}
The equations (\ref{motion-x})-(\ref{motion-J.3}) can be obtained according the rule $\dot z=\{z, H_2\}_D$. From
(\ref{Hamiltonian.0.02}) we conclude that our model describe spinning particle without gravimagnetic moment. In the
Hamiltonian formulation, equations of motion with gravimagnetic moment $\kappa$ have been proposed by Khriplovich
\cite{Khriplovich1989, Pomeranskii} adding non minimal interaction
$\frac{\lambda_1}{2}\frac{\kappa}{16}R_{\mu\nu\alpha\beta}S^{\mu\nu}S^{\alpha\beta}$ to the expression for $H_2$. It
would be interesting to find the corresponding Lagrangian formulation of the model.

Similarly to the spinless particle, we can exclude momenta $P^\mu$ from the Hamiltonian equations by using the
mass-shell condition. This yields second-order equation for the particle's position $x^\mu(\tau)$ (so we refer the
resulting equations as Lagrangian form of MPTD-equations). To achieve this, we observe that the equation
(\ref{motion-x}) is linear on $P$
\begin{equation}\label{xTP}
\dot x^\mu = \lambda_1  T^\mu_{\ \nu} P^\nu, \quad \textrm{with} \quad   T^\mu_{\ \nu}= \delta^\mu_\nu +
\frac{1}{8M^2c^2}S^{\mu\alpha}\theta_{\alpha \nu} \, .
\end{equation}
Using the identity
\begin{eqnarray}\label{Id0.0}
(S\theta S)^{\mu\nu}=-\frac{1}{2} (S\theta)S^{\mu\nu}, \quad \mbox{where} \quad
S\theta=S^{\alpha\beta}\theta_{\alpha\beta},
\end{eqnarray}
we find inverse of the matrix T
\begin{equation}\label{T-matrix}
\tilde T^\mu_{\ \ \nu} = \delta^\mu_{ \ \nu} - \frac{1}{8m^2c^2} S^{\mu\sigma}\theta_{\sigma\nu} \,, \qquad T\tilde T=1
\,,
\end{equation}
%
so (\ref{xTP}) can be solved with respect to $P^\mu$, $P^\mu=\frac{1}{\lambda_1}\tilde T^\mu{}_\nu\dot x^\nu$. We
substitute $P^\mu$ into the constraint $P^2+m^2c^2=0$, this gives expression for $\lambda_1$
\begin{equation}\label{L0}
\lambda_1= \frac{\sqrt {- G_{\mu\nu} \dot x^\mu \dot x^\nu}}{mc} \equiv \frac{\sqrt{-\dot x G \dot x}}{mc} \,.
\end{equation}
We have introduced the effective metric
\begin{equation}\label{G-metric}
G_{\mu\nu} \equiv \tilde T^\alpha_{\  \mu} g_{\alpha\beta}  \tilde T^{\beta}_{\ \nu} \, .
\end{equation}
From (\ref{xTP}) and (\ref{L0}) we obtain expression for $P_\mu$
\begin{eqnarray}\label{PTx}
P^\mu =\frac{mc}{\sqrt{-\dot x G \dot x}} \tilde T^\mu_{\ \ \nu}\dot x^\nu
=\frac{mc}{\sqrt{-\dot x G \dot x}}\left[ \dot x^\mu - \frac{1}{8m^2c^2} S^{\mu\nu}\theta_{\nu\sigma}\dot x^\sigma \right] \, ,
\end{eqnarray}
and Lagrangian form of the Tulczyjew condition
\begin{equation}\label{condition1a}
S^{\mu\nu} P_{\nu}=S^{\mu\nu}\tilde T_{\nu \sigma}\dot x^\sigma=0 \, .
\end{equation}
Using  the equations (\ref{PTx}) and (\ref{condition1a}) in (\ref{motion-P-2.2}) and (\ref{motion-J.3}) we finally
obtain
\begin{eqnarray}
\nabla\left[ \frac{ \tilde T^\mu_{\ \ \nu} \dot x^\nu}{\sqrt{-\dot xG\dot x}} \right] =- \frac{1}{4mc} R^\mu{}_{
\nu\alpha\beta}S^{\alpha\beta}\dot x^\nu \, , \label{byM13} \\
\nabla S^{\mu\nu}= \frac{1}{4mc\sqrt{-\dot xG\dot x}}\dot x^{[\mu} S^{\nu]\sigma} \theta_{\sigma\alpha}\dot x^\alpha
\, . \label{motionJ-5}
\end{eqnarray}
These equations, together with the conditions  (\ref{condition1a}) and (\ref{condition2}), form closed system for the
set ($x^\mu, S^{\mu\nu}$). The consistency of the constraints (\ref{condition1a}) and (\ref{condition2}) with the
dynamical equations is guaranteed by Dirac procedure for singular systems.

The Lagrangian considered above yields the fixed value of spin, that is this corresponds to an elementary particle. Let
us present the modification which leads to the theory with unfixed spin, and, similarly to Hanson-Regge approach
\cite{hanson1974}, with a mass-spin trajectory constraint. Consider the following Lagrangian
\begin{eqnarray}\label{aaa1}
L =-\frac{mc}{\sqrt{2}}\sqrt{-\dot x N \dot x - l^2\frac{\nabla\omega N \nabla\omega}{\omega^2} +\sqrt{\left[\dot x
N\dot x + l^2\frac{\nabla\omega N \nabla\omega}{\omega^2}\right]^2- 4l^2\frac{(\dot x N \nabla\omega )^2}{\omega^2}}},
\end{eqnarray}
where $l$ is a parameter with the dimension of length. Applying the Dirac procedure as in Section \ref{curvedS}, we
obtain the Hamiltonian
\begin{equation}\label{aaa2}
H=\frac{\lambda_1}{2}\left( P^2 + m^2 c^2 + \frac{\pi^2\omega^2}{l^2} \right) + \lambda_5 (\omega\pi) +\lambda_6
(P\omega) +\lambda_7 (P\pi )\,,
\end{equation}
which turns out to be combination of the first-class constraints $P^2 + m^2 c^2 + \frac{\pi^2\omega^2}{l^2}=0$,
$\omega\pi=0$ and the second-class constraints $P\omega=0$, $P\pi=0$. The Dirac procedure stops on the first stage,
that is there are no of secondary constraints. As compared with (\ref{L-curved}), the first-class constraint
$\pi^2-\frac{\alpha}{\omega^2}=0$ does not appear in the present model. Due to this, square of spin is not fixed,
$S^2=8(\omega^2\pi^2-\omega\pi)\approx8\omega^2\pi^2$. Using this equality, the mass-shell constraint acquires the
string-like form
\begin{equation}\label{aaa3}
P^2 + m^2 c^2 + \frac{1}{8l^2}S^2=0.
\end{equation}
The model has four physical degrees of freedom in the spin-sector. As the independent gauge-invariant degrees of
freedom, we can take three components $S^{ij}$ of the spin-tensor together with any one product of conjugate
coordinates, for instance, $\omega^0\pi^0$.

\section{MPTD equations and dynamics of representative point of a rotating body}\label{Repr}

In this section we discuss MPTD-equations of a rotating body in the form studied by Dixon (for the relation of the
Dixon equations with those of Papapetrou and Tulczyjew see p. 335 in \cite{Dixon1964})
\begin{eqnarray}
\nabla P^\mu=-\frac 14 R^\mu{}_{\nu\alpha\beta}S^{\alpha\beta}\dot x^\nu \equiv-\frac 14 (\theta\dot x)^\mu \, ,\label{r1}\\
\nabla S^{\mu\nu}= 2(P^\mu \dot x^\nu - P^\nu \dot x^\mu)\, , \label{r2} \\
S^{\mu\nu}P_\nu  =0, \label{r3}
\end{eqnarray}
and compare them with equations of motion of our spinning particle. In particular, we show that the effective metric
$G_{\mu\nu}$ also emerges in this formalism. MPTD-equations appeared in multipole approach to description of a body
\cite{Mathisson:1937zz, Papapetrou:1951pa, Tulc, Dixon1964, Dixon1965, Trautman2002, Costa2012, Costa2014}, where the
energy-momentum of the body is modelled by a set of multipoles. In this approach $x^\mu(\tau)$ is called representative
point of the body, we take it in arbitrary parametrization $\tau$ (contrary to Dixon, we do not assume the proper-time
parametrization, that is we do not add the equation $g_{\mu\nu}\dot x^\mu\dot x^\nu=-c^2$ to the system above).
$S^{\mu\nu}(\tau)$ is associated with inner angular momentum, and $P^\mu(\tau)$ is called momentum. The first-order
equations (\ref{r1}) and (\ref{r2}) appear in the pole-dipole approximation, while the algebraic equation (\ref{r3})
has been added by hand\footnote{For geometric interpretation of the spin supplementary condition in the multipole
approach see \cite{Costa2014}.}. After that, the number of equations coincides with the number of variables.

To compare MPTD-equations with those of previous section, we first observe some useful consequences of the system
(\ref{r1})-(\ref{r3}).

Take derivative of the constraint, $\nabla(S^{\mu\nu}P_\nu)=0$, and use (\ref{r1}) and (\ref{r2}), this gives the
expression
\begin{eqnarray}\label{r4}
(P\dot x)P^\mu=P^2\dot x^\mu+\frac18(S\theta \dot x)^\mu,
\end{eqnarray}
which can be written in the form
\begin{eqnarray}\label{r5}
P^\mu=\frac{P^2}{(P\dot x)}\left(\delta^\mu{}_\nu+\frac{1}{8P^2}(S\theta)^\mu{}_\nu\right)\dot x^\nu
\equiv\frac{P^2}{(P\dot x)}\tilde{\mathcal{T}}^\mu{}_\nu\dot x^\nu.
\end{eqnarray}
Contract (\ref{r4}) with $\dot x_\mu$. Taking into account that $(P\dot x)<0$, this gives
$(P\dot x)=-\sqrt{-P^2}\sqrt{-\dot x\tilde{\mathcal{T}}\dot x}$.
Using this in Eq. (\ref{r5}) we obtain
\begin{eqnarray}\label{r7}
P^\mu=\frac{\sqrt{-P^2}}{\sqrt{-\dot x\tilde T\dot x}}(\tilde T\dot x)^\mu, \qquad \tilde
{\mathcal{T}}^\mu{}_\nu=\delta^\mu{}_\nu+\frac{1}{8P^2}(S\theta)^\mu{}_\nu.
\end{eqnarray}
For the latter use we observe that in our model with composite $S^{\mu\nu}$ we used the identity (\ref{Id0.0}) to
invert $T$, then the Hamiltonian equation (\ref{motion-x}) has been written in the form (\ref{PTx}), the latter can be
compared with (\ref{r7}).

Contracting  (\ref{r2}) with $S_{\mu\nu}$ and using  (\ref{r3}) we obtain $\frac{d}{d\tau}(S^{\mu\nu}S_{\mu\nu})=0$,
that is, square of spin is a constant of motion. Contraction of  (\ref{r4}) with $P_\mu$  gives $(PS\theta\dot x)=0$.
Contraction of (\ref{r4}) with $(\dot x\theta)_\mu$ gives $(P\theta\dot x)=0$. Contraction of (\ref{r1}) with $P_\mu$,
gives $\frac{d}{d\tau}(P^2)=-\frac12(P\theta\dot x)=0$, that is $P^2$ is one more constant of motion, say $k$,
$\sqrt{-P^2}=k=\mbox{const}$ (in our model this is fixed as $k=mc$). Substituting  (\ref{r7}) into the equations
(\ref{r1})-(\ref{r3}) we now can exclude $P^\mu$ from these equations, modulo to the constant of motion
$k=\sqrt{-P^2}$.

Thus, square of momentum can not be excluded from the system (\ref{r1})-(\ref{r4}), that is MPTD-equations
in this form do not represent a Hamiltonian system for the pair $x^\mu, P^\mu$.
To improve this point, we note that Eq. (\ref{r7}) acquires a conventional form (as the expression for conjugate
momenta of $x^\mu$ in the Hamiltonian formalism), if we add to the system (\ref{r1})-(\ref{r3}) one more equation,
which fixes the remaining quantity $P^2$ (Dixon noticed this for the body in electromagnetic field, see his eq. (4.5)
in \cite{Dixon1965}). To see, how the equation could look, we note that for non-rotating body (pole approximation) we
expect equations of motion of spinless particle, $\nabla p^\mu=0$, $p^\mu=\frac{mc}{\sqrt{-\dot xg\dot x}}\dot x^\mu$,
$p^2+(mc)^2=0$. Independent equations of the system (\ref{r1})-(\ref{r4}) in this limit read $\nabla P^\mu=0$,
$P^\mu=\frac{\sqrt{-P^2}}{\sqrt{-\dot xg\dot x}}\dot x^\mu$. Comparing the two systems, we see that the missing
equation is the mass-shell condition $P^2+(mc)^2=0$. Returning to the pole-dipole approximation, an admissible equation
should be $P^2+(mc)^2+f(S, \ldots)=0$, where $f$ must be a constant of motion. Since the only constant of motion in
arbitrary background is $S^2$, we have finally
\begin{eqnarray}\label{r9.1}
P^2=-(mc)^2-f(S^2).
\end{eqnarray}
With this value of $P^2$, we can exclude $P^\mu$ from MPTD-equations, obtaining closed system with second-order
equation for $x^\mu$. We substitute (\ref{r7}) into (\ref{r1})-(\ref{r3}), this gives
\begin{eqnarray}
\nabla\frac{(\tilde{\mathcal{T}}\dot x)^\mu}{\sqrt{-\dot x\tilde{\mathcal{T}}\dot
x}}=-\frac{1}{4\sqrt{-P^2}}(\theta\dot x)^\mu,
\qquad \qquad \label{r9} \\
\nabla S^{\mu\nu}=
-\frac{1}{4\sqrt{-P^2}{\sqrt{-\dot x\tilde{\mathcal{T}}\dot x}}}\dot x^{[\mu}(S\theta\dot x)^{\nu]}, \label{r10} \\
(SS\theta\dot x)^\mu=-8P^2(S\dot x)^\mu, \qquad \qquad \qquad \label{r11}
\end{eqnarray}
where (\ref{r9.1}) is implied. They determine evolution of $x^\mu$ and $S^{\mu\nu}$ for each given function $f(S^2)$.

It is convenient to introduce the effective metric ${\mathcal{G}}$ composed from the "tetrad field"
$\tilde{\mathcal{T}}$
\begin{equation}\label{r12}
{\mathcal{G}}_{\mu\nu} \equiv g_{\alpha\beta} \tilde{\mathcal{T}}^{\alpha}{}_{\mu} \tilde{\mathcal{T}}^{\beta}{}_{\nu}
\, .
\end{equation}
Eq. (\ref{r11}) implies the identity
\begin{eqnarray}\label{r13}
\dot x\tilde{\mathcal{T}}\dot x=\dot x{\mathcal{G}}\dot x,
\end{eqnarray}
so we can replace $\sqrt{-\dot x\tilde{\mathcal{T}}\dot x}$ in (\ref{r9})-(\ref{r11}) by $\sqrt{-\dot
x{\mathcal{G}}\dot x}$.

In resume, we have presented MPTD-equations in the form
\begin{eqnarray}
P^\mu=\frac{\sqrt{-P^2}}{\sqrt{-\dot x{\mathcal{G}}\dot x}}(\tilde{\mathcal{T}}\dot x)^\mu \, , \quad  \nabla
P^\mu=-\frac 14 (\theta\dot x)^\mu \, , \quad \nabla S^{\mu\nu}= 2P^{[\mu} \dot x^{\nu]} \, , \quad S^{\mu\nu}P_\nu  =0
\, , \label{r003} \\
P^2+(mc)^2+f(S^2)=0 \, , \qquad \qquad \qquad \qquad \qquad \qquad  \label{r004} \\
S^2 \quad \mbox{is a constant of motion} \, , \qquad \qquad \qquad \qquad \qquad \quad  \label{r005}
\end{eqnarray}
with $\tilde{\mathcal{T}}$ given in (\ref{r7}). Now we are ready to compare them with Hamiltonian equations of our
spinning particle, which we write here in the form
\begin{eqnarray}
P^\mu =\frac{mc}{\sqrt{-\dot x G \dot x}}(\tilde T\dot x)^\mu, \quad \nabla P^\mu =-\frac 14 (\theta\dot x)^\mu \, ,
\quad \nabla S^{\mu\nu} = 2P^{[\mu} \dot x^{\nu]}\, , \quad
S^{\mu\nu}P_\nu  =0 \, , \label{m003} \\
P^2+(mc)^2=0 \, , \qquad \qquad \qquad \qquad \qquad \qquad \qquad \label{m004} \\
S^2=8\alpha \, , \qquad \qquad \qquad \qquad \qquad  \qquad \qquad \qquad \label{m005}
\end{eqnarray}
with $\tilde T$ given in (\ref{T-matrix}). Comparing the systems, we see that our spinning particle has fixed values of
spin and canonical momentum, while for MPTD-particle the spin is a constant of motion and momentum is a function of
spin. We conclude that all the trajectories of a body with given $m$ and $S^2=\beta$ are described by our spinning
particle with spin $\alpha=\frac{\beta}{8}$ and with the mass equal to $\sqrt{m^2-\frac{f^2(\beta)}{c^2}}$. In this
sense our spinning particle is equivalent to MPTD-particle\footnote{We point out that our final conclusion remains true
even we do not add (\ref{r9.1}) to MPTD-equations: to study the class of trajectories of a body with $\sqrt{-P^2}=k$
and $S^2=\beta$ we take our spinning particle with $m=\frac{k}{c}$ and $\alpha=\frac{\beta}{8}$.}.

MPTD-equations in the Lagrangian form (\ref{r9})-(\ref{r11}) can be compared with
(\ref{condition1a})-(\ref{motionJ-5}).

\section{Lagrangian form of MPTD-equations}\label{Lf}
Here we shortly discuss some immediate consequences which can be obtained from the Lagrangian form
(\ref{condition1a})-(\ref{motionJ-5}), (\ref{condition2}) of MPTD-equations.

In the spinless limit the equation (\ref{byM13}) turns into the geodesic equation. Spin causes deviations from the
geodesic motion due to right hand side of this equation, as well as due to presence of the tetrad field $\tilde T$ and
the effective metric $G$ in the left hand side. In the Newtonian limit  the original metric $ g_{\mu\nu}(x) $ can be
presented through the Newton potential in which a test body is immersed. The presence of the $ G_{\mu\nu} $ could be
thought as a contribution to this potential when spin of the body is taken in account. Let us compute manifest form of
$G$ in the field with nearly flat metric
\begin{equation}
g_{\mu\nu}=\eta_{\nu\mu} + h_{\mu\nu}, \quad |h_{\mu\nu}| \ll 1 \, .
\end{equation}
To linear order in $h_{\mu\nu}$ the curvature tensor is $R^{(1)}_{\mu\nu\alpha\beta} = \frac 12 ( h_{\mu\beta,
\nu\alpha} +  h_{\nu\alpha, \mu\beta} -h_{\nu\beta , \mu\alpha} - h_{\mu\alpha , \nu\beta} )$, hence
$\theta^{(1)}_{\mu\nu}= R^{(1)}_{\mu\nu\alpha\beta} S^{\alpha\beta} = (h_{\mu\alpha,\beta\nu} -
h_{\nu\alpha,\beta\mu})S^{\beta\alpha}$, where the comma denotes partial derivative. The effective metric in
the weak field approximation reads
\begin{equation}\label{Newton1}
G^{(1)}_{\mu\nu} = g_{\mu\nu} - \frac{1}{8m^2c^2} \left( \eta_{\mu\alpha} S^{\alpha\beta} \theta^{(1)}_{\beta \nu} +
\eta_{\nu\alpha} S^{\alpha\beta} \theta^{(1)}_{\beta\mu} \right).
\end{equation}
Let us consider the Newtonian solution to the linearized Einstein equations
\begin{equation}\label{Newton}
h_{00}=-2\phi, \quad h_{ij}= -2\delta_{ij} \phi , \quad h_{\mu 0}=0\, ,
\end{equation}
with $\phi=-\frac{k}{r}$. Using the three-dimensional spin-vector and the dipole electric moment (\ref{FF.2}), the
time-time component of the effective metric is
\begin{eqnarray}\label{Newton2}
G_{00} = -1 + \frac{2k}{r} + \frac{k}{2m^2c^2r^3}\left[3({\bf D}  \cdot {\bf n})^2  - {\bf D}^2 \right] \, ,
\end{eqnarray}
where ${\bf n}={\bf r}/r$. Contrary to the Newtonian solution (\ref{Newton}) the space-time components of $G_{\mu\nu}$
are different to zero
\begin{equation}\label{Newton3}
G_{i0}= \frac{3k}{4m^2c^2 r^3} \left[({\bf D\times s})_i -2({\bf D\cdot n}) ({\bf n\times s})_i -n_i ({\bf D\times
s)\cdot n} \right] \, .
\end{equation}
For the space-space components we found
\begin{eqnarray}\label{Newton4}
G_{ij} = \delta_{ij} +\frac{2k}{r} \delta_{ij} + \frac{k}{2m^2 c^2 r^3} \left\{ \left[3\hat n_i \hat n_j -
5\delta_{ij}  \right] {\bf s}^2 -5s_i s_j  + D_i D_j \right. \nonumber \\
\left.  -\frac 32 \left[ ({\bf s\cdot n}) s_{(i}n_{j)} + ({\bf D \cdot n}) D_{(i} n_{j)}  \right]
 -12 ({\bf n \times s})_i ({\bf n \times s})_j  \right\} \,.
\end{eqnarray}
We point out that the expressions (\ref{Newton1})-(\ref{Newton4}) are written without any approximation with respect to
spin.
The contributions due to spin over long distances will be very small, then in the Newtonian limit a spinning particle
behaves almost as a spinless one.  Probably at short distances the contributions may be important; to verify this,
other geometries should be considered.

Our formulation reveals one more novel property of MPTD-equations: mean position of a rotating body will be represented
by non-commutative operators in quantum theory. Indeed, to construct the quantum theory of a system with second-class
constraints, one should pass from Poisson to Dirac bracket \cite{13, gitman1990quantization, deriglazov2010classical}.
Then one look for operators of basic variables with commutators resembling the Dirac bracket. For our case the Dirac
bracket is given by (\ref{DB}). This yields highly non-commutative algebra for the position variables
\begin{equation}\label{NC}
\{ x^\mu , x^\nu\}_D =\frac{2 \omega^{[\mu} \pi^{\nu]} }{M^2c^2}\equiv\frac{ S^{\mu \nu} }{M^2c^2} \,.
\end{equation}
In the result, the position space is endowed with noncommutative structure which originates from accounting of spin
degrees of freedom. We point out that non relativistic spinning particle implies canonical algebra of position
operators, see \cite{deriglazovMPL2010, DPM1}. So the deformation (\ref{NC}) arises as a relativistic correction
induced by spin. It is known that formalism of dynamical systems with second-class constraints implies a natural
possibility to incorporate noncommutative geometry into the framework of classical and quantum theory \cite{hanson1974,
AAD10, AAD11, Abreu:2010mt, Amorim:2010qj}. Our model represents an example where physically interesting noncommutative
particle (\ref{NC}) emerges in this way. For the case, the "parameter of non-commutativity" is proportional to
spin-tensor. This allowed us \cite{DPW2} to explain contradictory results concerning first relativistic corrections due
to spin obtained by different authors.

Consider the background metric which admits the Killing vector $\xi_\mu$, $\xi_{\mu;\nu} +\xi_{\nu;\mu}=0$ (the
semicolon means the covariant derivative). Then the infinitesimal transformation
\begin{equation}\label{isometry}
x'^\mu = x^\mu + \varepsilon \xi^\mu(x), \quad \varepsilon<<1 \, .
\end{equation}
generates isometry of the metric, that is leaves it form-invariant, $g'_{\mu\nu}(y)=g_{\mu\nu}(y)$. For the spinless
particle the isometry generates the conserved quantity $\frac{\partial L}{\partial\dot x^\mu}\xi^\mu$. A natural
question is, whether this remains true for a vector model of spin, where the particle do not follows a geodesic
trajectory? From the transformation law of $\omega^\mu$
\begin{equation}
\omega'^\mu(\tau)=\frac{\partial x'^\mu}{\partial x^\alpha} \omega^\alpha(\tau) = \left(\delta^\mu_\alpha +\varepsilon
\xi^\mu_{ \  ,\alpha}\right) \omega^\alpha(\tau),
\end{equation}
we deduce  that
%
$\delta \omega^\mu = \omega^\mu(\tau) - \omega^\mu(\tau)= \varepsilon \omega^\nu \xi^\mu_{\ \ ,\nu}$,
%
which corresponds to the transformation law of a form-invariant vector field. By the Noether's theorem the quantity
\begin{equation}\label{Noether1}
J^{(\xi)}= \frac{\partial L}{\partial\dot x^\mu} \delta x^\mu + \frac{\partial L}{\partial \dot\omega^\mu}
\delta\omega^\mu= p_\mu \xi^\mu +\xi^{\mu}_{\ \ ,\nu}  \pi_\mu \omega^\nu \, ,
\end{equation}
is conserved. In terms of $S^{\mu\nu}$ and $P_\mu$ this coincides with that of \cite{Hojman3},
%
$J^{(\xi)}=P^\mu\xi_\mu - \frac{1}{4}S^{\mu\nu}\xi_{\mu;\nu}$.
%
Using equations (\ref{motion-P-2.2}) and (\ref{motion-J.3}), it is easy to confirm that $J^{(\xi)}$ is conserved. We
conclude that an isometry of spinless particle remains the isometry for the vector models of spin. However the
conserved quantity acquires the spin-dependent term $-\frac14 S^{\mu\nu}\xi_{\mu;\nu}$.


\section{Conclusions}\label{Discussion}
In this work we have presented the Lagrangian action without auxiliary variables (\ref{L-curved})  for  description of
spinning particle in an arbitrary curved background.  The supplementary spin conditions (\ref{condition1}) and
(\ref{condition2}) are guaranteed by the set of constraints (\ref{primary}) and (\ref{secondary}) arising from our
singular Lagrangian in the Hamiltonian formalism. Due to this, spin has two physical degrees of freedom, as it should
be for spin one-half particle. Besides, the reparametrization invariance of the action generates the mass-shell
constraint $P^2+(mc)^2=0$. The description of spin on the base of vector-like variable allows us to construct also the
Lagrangian (\ref{aaa1}) with unfixed value of spin and string-like mass-shell constraint (\ref{aaa3}), as in the
Hanson-Regge model of relativistic top. In the model (\ref{aaa1}) appeared the fundamental length scale and spin has
four physical degrees of freedom.

We showed that our spinning particle can be used to study dynamics of a rotating body in curved background: all the
trajectories of MPTD-equations with given values of integration constants, $\sqrt{-P^2}=k$ and $S^2=\beta$, are
described by our spinning particle with $m=\frac{k}{c}$ and $\alpha=\frac{\beta}{8}$. In this sense the expression
(\ref{L-curved}) yields the Lagrangian formulation  of MPTD-equations, the latter correspond to minimal interaction of
the particle with gravity. This demonstrates the effectiveness of classical description of spin on the base of
vector-like non Grassmann variable. We have explored our formulation to obtain, in unambiguous way, the closed system
of equations (\ref{condition1a})-(\ref{motionJ-5}), (\ref{condition2}) for the set $x^\mu , S^{\mu\nu}$. Some immediate
consequences of this form of  MPTD-equations have been discussed in Section \ref{Lf}. In particular, in the Lagrangian
form of MPTD-equations, instead of the original metric $g_{\mu\nu}$ emerges the effective metric
$G_{\mu\nu}=g_{\mu\nu}+H_{\mu\nu}$ with spin and field-dependent contribution $H_{\mu\nu}$. According to
(\ref{G-metric}), the matrix (\ref{T-matrix}), which links canonical momentum and velocity, plays the role of a tetrad
field to compose the effective metric. The effective metric determines behavior of MPTD-particle in ultra-relativistic
limit \cite{deriglazovMPL2015}.

\section*{Acknowledgments}
The work of AAD has been supported by the Brazilian foundations CNPq (Conselho Nacional de Desenvolvimento Científico e
Tecnológico - Brasil) and FAPEMIG (Fundação de Amparo à Pesquisa do estado de Minas Gerais- Brasil). WGR thanks CAPES
for the financial support (Program PNPD/2011).

\end{document}